\documentclass[aps,prl,twocolumn,floats,superscriptaddress]{revtex4-1}

\usepackage{graphicx}
\usepackage{multirow}
\usepackage{amsmath}
\usepackage{amssymb}
\usepackage{amsfonts}
\usepackage{soul}
\usepackage{setspace}
\usepackage{float}
\usepackage{pdfpages}
\usepackage{color}
\usepackage{colortbl,array}
\definecolor{nblack}{rgb}{0,0,0}
\definecolor{nblue}{rgb}{0.2,0.2,0.7}
\definecolor{nred}{rgb}{0,0,0}
\definecolor{ngreen}{rgb}{0.2,0.6,0.2}
\definecolor{nbluedark}{rgb}{0.15,0.15,0.6}
\definecolor{a}{rgb}{0.7,0.2,0.2}
\definecolor{b}{rgb}{0,0,0}
\definecolor{c}{rgb}{0.2,0.6,0.2} 

\newcommand{\white}{\color{white}}
\newcommand{\blk}{\color{nblack}}

\makeatletter
\def\blfootnote{\xdef\@thefnmark{}\@footnotetext}
\makeatother

\usepackage{soul}
\providecommand{\openone}{\leavevmode\hbox{\small1\kern-3.8pt\normalsize1}}

\newcommand{\uibke}{Institut f\"ur Experimentalphysik, Universit\"at Innsbruck, Technikerstrasse 25, 6020 Innsbruck, Austria}
\newcommand{\gisin}{Group of Applied Physics, University of Geneva, Geneva, Switzerland}
\newcommand{\iqoqi}{Institut f\"ur Quantenoptik und Quanteninformation, \"Osterreichische Akademie der Wissenschaften,Technikerstrasse 21A, 6020 Innsbruck, Austria\\ $^*$ These authors contributed equally to this work.}

\begin{document}
\title{Device-independent demonstration of genuine multipartite entanglement}

\author{Julio~T.~Barreiro$^{*a}$\footnote{Present address: Fakult\"at f\"ur Physik, Ludwig-Maximilians-Universit\"at M\"unchen \& Max-Planck Institute of Quantum Optics, Germany}}
\affiliation{\uibke}
\author{Jean-Daniel~Bancal$^{*b}$\footnote{Present address: Centre for Quantum Technologies, National University of Singapore, Singapore}}
\affiliation{\gisin}
\author{Philipp~Schindler}
\affiliation{\uibke}
\author{Daniel~Nigg}
\affiliation{\uibke}
\author{Markus~Hennrich}
\affiliation{\uibke}
\author{Thomas~Monz}
\affiliation{\uibke}
\author{Nicolas~Gisin}
\affiliation{\gisin}
\author{Rainer~Blatt}
\affiliation{\uibke}
\affiliation{\iqoqi}

\begin{abstract}
Entanglement in a quantum system can be demonstrated
  experimentally by performing the measurements prescribed by an
  appropriate entanglement
  witness~\cite{guhne-physrep-474-1}. However, the unavoidable
  mismatch between the implementation of measurements in practical
  devices and their precise theoretical modelling generally results in
  the undesired possibility of false-positive entanglement
  detection~\cite{rosset-arxiv-1203.0911}. Such scenarios can be
  avoided by using the recently developed device-independent
  entanglement witnesses (DIEWs) for genuine multipartite
  entanglement~\cite{bancal-prl-106-250404}. Similarly to Bell
  inequalities, DIEWs only assume that consistent measurements are
  performed locally on each subsystem. No precise description of the
  measurement devices is required. Here we report an experimental test
  of DIEWs on up to six entangled $^{40}$Ca$^{+}$ ions. We also
  demonstrate genuine multipartite quantum nonlocality between up to
  six parties with the detection loophole closed.
\footnote{Present address: Fakult\"at f\"ur Physik, Ludwig-Maximilians-Universit\"at M\"unchen \& Max-Planck Institute of Quantum Optics, Germany}\hspace{-0.31cm}
\footnote{Present address: Centre for Quantum Technologies, National University of Singapore, Singapore}\hspace{-0.31cm}
{\white$^\blacksquare$}
\end{abstract}
\maketitle

Entanglement enables many quantum tasks, including scalable quantum
communication~\cite{briegel-prl-81-5932}, secure quantum key
distribution~\cite{ekert-prl-67-661,gisin-rmp-74-145} and quantum
computing~\cite{briegel-natphys-5-19}. Entanglement thus represents a
key resource for quantum information processing. As the experimental
systems developed to host and manipulate entangled states
diversify~\cite{dicarlo-nature-467-574,neumann-science-320-1326,saffman-rmp-82-2313,blatt-nature-453-1008},
clear tests are desired that are able to certify properly the
entanglement of such states. In particular, tests of genuine
multipartite entanglement are needed to show that the entanglement
truly involves all of their constituents~\cite{guhne-physrep-474-1}.

Tools to detect genuine multipartite entanglement have been
developed~\cite{guhne-physrep-474-1} and used to demonstrate, for
instance, entanglement between up to 8-10 photonic
qubits~\cite{yao-natphot-6-225,gao-natphys-6-331} or 14
ions~\cite{monz-prl-106-130506}. Namely, it was shown that every
multipartite entangled state can be detected by a hermitian operator
known as an entanglement witness, whose expectation value is negative
on the considered state but positive on all biseparable states. Any
measurement of a witness yielding a negative value thus certifies that
the measured state is multipartite entangled. Experimentally speaking,
entanglement witnesses are convenient because they can be expressed in
terms of simple measurable observables specific to the physical
system, in contrast to joint measurements over multiple copies.

Although entanglement witnesses are very well understood from a
theoretical point of view, their application to practical systems
strongly relies on the way that concrete systems are modeled. For
example, entanglement witnesses generally take different forms
depending on the dimension of the Hilbert space under
consideration. An artificial truncation of this space can thus lead to
wrong entanglement detection (e.g., in photon
number~\cite{semenov-pra-83-032119}). Similarly, any mismatch between
the measurements prescribed and those performed, such as systematic
errors, can also lead to flawed entanglement
detection~\cite{rosset-arxiv-1203.0911}.  Entanglement witnesses can
also be evaluated on a state reconstructed by
tomography~\cite{altepeter-lnp-649-113,lvovsky-rmp-81-299}, but the
reconstruction also relies on a detailed model of the measurement
devices and Hilbert space dimensionality (see
e.g. Ref.~\cite{audenaert-njp-11-023028} for possible consequences of
using a wrong model). In particular, quantum state tomography is also
sensitive to systematic measurement
errors~\cite{rosset-arxiv-1203.0911}.

Systematic errors and untenable assumptions may lead one to conclude
that any proper experimental entanglement certification requires a
particularly good prior knowledge of the measurement devices. But this
is not the case: a rudimentary measurement model, whose validity can
in principle be checked in practice, is sufficient to demonstrate
entanglement properties of quantum systems. Namely, entanglement can
be certified under the sole hypothesis that each subsystem can be
addressed individually with independent measurement devices. Indeed,
this condition allows one to ensure that, upon measurement of a
separable quantum state, the raw statistics satisfy all Bell
inequalities~\cite{guhne-physrep-474-1}. Violation of a Bell
inequality under these conditions thus certifies that the measured
state was entangled. In contrast with conventional entanglement
witnesses, this conclusion holds independently of the internal
mechanisms of the measurement devices used in the
experiment~\cite{acin-prl-97-120405} (device independently as in
recent tests of classical and quantum
dimensions~\cite{gallego-prl-105-230501,hendrych-natphys-8-588,ahrens-natphys-8-592}). Because
Bell inequalities, in general, do not detect genuine multipartite
entanglement, dedicated device-independent witnesses were recently
developed in the form of
DIEWs~\cite{bancal-prl-106-250404,pal-pra-83-062123,bancal-jphysa-45-125301}.

Several architectures have realized multi-particle
entanglement~\cite{dicarlo-nature-467-574,neumann-science-320-1326,saffman-rmp-82-2313,blatt-nature-453-1008}
but trapped ions have demonstrated particularly high fidelity in the
preparation, manipulation and detection of quantum states.  Similar to
other systems, however, such ion trap architectures are affected by
crosstalk.  This imperfection occurs on systems in which measurements
or operations intended on individual qubits affect neighboring qubits
(close physically or in frequency, for example). Because crosstalks
were not considered in the original derivation of
DIEWs~\cite{bancal-prl-106-250404}, we adapted the bound of the tested
witnesses to take this imperfection into account.  We then demonstrate
genuine multipartite entanglement between three, four and six trapped
ions with the aid of these DIEWs.

As described in detail in Ref.~\cite{bancal-prl-106-250404}, let us
consider an entangled state consisting of $n$ parties or subsystems
and a device-independent scenario where the measurement settings and
outcomes are only refered to by indices and without alluding to the
specifics of how these measurements are performed.  We consider that a
party $j$ of this state is measured in one of $m$ possible ways.  We
thus index this measurement setting by
 $s_j=0,\ldots,m-1$. We also consider that the local measurement of
each party yields one of two outcomes that we index by 
$r_j=0,1$. We can thus identify a measurement on the entire
$n$-partite system by the vector $\vec s=(s_1,\ldots,s_n)\in
\{0,\ldots,m-1\}^n$. Similarly, we describe each possible outcome of a
measurement on the entire system by a vector $\vec
r=(r_1,\ldots,r_n)\in\{0,1\}^n$.  After measuring the state, the
conditional probability, or raw statistics, of observing the outcome
$\vec r$ given that the choice of measurement settings $\vec s$ is
applied to each of the $n$ qubits is denoted by $P(\vec r|\vec
s)$. For an $n$-partite state with the described measurements and
outcomes, we define the DIEW parameter as
\begin{eqnarray}
I_{nm} = \sum_{s\equiv0\;(\text{mod}\; m)} (-1)^{s/m}E_{\vec{s}}+\nonumber\\
          \sum_{s\equiv1\;(\text{mod}\; m)} (-1)^{(s-1)/m}E_{\vec{s}}\label{eq:diew}
\end{eqnarray}
where $E_{\vec{s}}=\sum_{\vec{r}}(-1)^r P(\vec{r}|\vec{s})$ is the
$n$-partite correlator; $s=\sum_j s_j$ and $r=\sum_j r_j$
(see Ref.~\cite{bancal-jphysa-45-125301}).

As shown in Ref.~\cite{bancal-jphysa-45-125301}, because all
biseparable quantum correlations satify
$$I_{nm}\le 2 m^{n-2}\cot{(\pi/2 m)}\equiv B_{nm},$$ this inequality
can be used as a DIEW to detect \emph{genuine $n$-partite
  entanglement}. In contrast with usual entanglement witnesses,
$I_{nm}$ provides no explicit description of the observables to be
measured on each subsystem.  Yet, every practical evaluation of
Eq.~\eqref{eq:diew} is necessarily performed with some measurement
operators $M_{r_j|s_j}^j$ so that the observed correlations can be
written as $P(\vec r|\vec{s})=\text{Tr}\left[\overset{n}{\underset{j=1}{\bigotimes}}M_{r_j|s_j}^j\rho\right]$.
Here $M_{r_j|s_j}^j$ is the measurement operator acting locally on
subsystem $j$ corresponding to setting $s_j$ and outcome $r_j$, and
$\rho$ is the measured quantum state. The DIEW parameter can thus be
understood as the expectation value of a quantum operator $\mathcal
I_{nm}$ whose biseparable bound is given by $B_{nm}$ (see
Supplementary Information for a definition of $\mathcal I_{nm}$) . In
particular, the quantity $\mathcal W_{nm} = B_{nm}\openone-\mathcal
I_{nm}$ is a standard entanglement witness for genuine multipartite
entanglement~\cite{guhne-physrep-474-1}. Therefore, any practical
evaluation of a DIEW amounts to test a standard witness that is
adapted to the measurement operators available during the experiment. {\color{nred}In addition, DIEWs need less measurements and data processing than complete state tomography (see Methods).}

We focus on a class of multipartite states routinely prepared in our
experiment, the genuinely entangled $n$-qubit state
$|GHZ\rangle_n=\frac{1}{\sqrt{2}}\left(|0\rangle^{\otimes
  n}+|1\rangle^{\otimes n}\right)$ which yields the maximum value of
our DIEW parameter
$$I_{nm}^{max}=2m^{n-1}\cos(\pi/2m)>B_{nm}$$ when the following $m$
measurements act on each qubit $j$ in the $x$-$y$ plane of the Bloch
sphere
\begin{eqnarray}
\cos{(\phi_{s_j})}\sigma_x^{(j)} + \sin{(\phi_{s_j})}\sigma_y^{(j)}\blk,\quad\text{with\ \ \ \ \ \ \ \ \ \ }\nonumber\\
\phi_{s_j}=-\frac{\pi}{2mn}+s_j\frac{\pi}{m}\quad\text{for}\quad s_j=0,\ldots,m-1,\ \ \ \ \label{eq:setts}
\end{eqnarray}
where $\sigma_{x,y,z}^{(j)}$ are the Pauli matrices for qubit $j$ (see
Supplementary Information for the optimization procedure). For a given
measurement $\vec{s}$ on all qubits, this measurement setting is then
equivalent to rotating each qubit with
$\exp(-i\frac{\phi_{s_j}}{2}\sigma_z^{(j)})$ for every $j$, followed
by a collective rotation of all qubits with
$\exp(-i\frac{\theta}{2}\sum_j\sigma_x^{(j)})$ where $\theta=\pi/2$,
and a measurement in the computational basis. Our witness is indeed
device-independent: if these operations are not implemented as
expected, because the $\phi_{s_j}$ take unprescribed values, or
$\theta\neq\pi/2$, and consequently the measurement operators
$M^j_{r_j|s_j}$ change, then by
construction~\cite{bancal-prl-106-250404} the bound $I_{nm}\leq
B_{nm}$ still holds for all biseparable states whatever the actual
measurement operators $M^j_{r_j|s_j}$ are, thus avoiding false
positive entanglement detection.

In the experiment each qubit was encoded in the internal electronic
Zeeman levels 4S$_{1/2}(m=-1/2)=|1\rangle$ and
3D$_{5/2}(m=-1/2)=|0\rangle$ of a $^{40}$Ca$^+$ ion. Three, four or
six ions were confined to a string by a linear Paul trap and cooled to
the ground state of the axial center-of-mass
mode~\cite{schmidt-kaler-apblo-77-789}. The entanglement among qubits
was realized through a
M\o{}lmer-S\o{}rensen~\cite{molmer-prl-82-1835,sackett-nat-404-256}
entangling operation using a bichromatic light field collectively
illuminating all ions~\cite{monz-prl-106-130506} [e.g.,
  Fig.~1a]. Single-qubit rotations
$\exp(-i\frac{\phi_j}{2}\sigma_z^{(j)})$ were driven by a far
off-resonantly detuned laser pulse focused on the target ion $j$ and
inducing an AC-Stark effect, where $\phi_j$ is determined by the
detuning and the pulse duration [e.g., Fig.~1b]. Collective qubit
rotations $\exp(-i\frac{\theta}{2}\sum_j\sigma_x^{(j)})$ were driven
by a laser pulse exciting the qubit transition, where $\theta$ is
given by the Rabi frequency of the qubits and the duration of the
pulse [e.g., Fig.~1c]. At the end the qubits were measured in the
computational basis through an electron-shelving technique scattering
light on the dipole transition $4S_{1/2}\leftrightarrow 4P_{1/2}$. The
scattered light was detected with a CCD camera that resolves each
ion's fluorescence.

\setcounter{figure}{0}
\renewcommand{\thefigure}{\textbf{\arabic{figure}}}
\renewcommand{\figurename}{\textbf{Figure}}
\begin{figure}[!h]
\includegraphics[width=86mm]{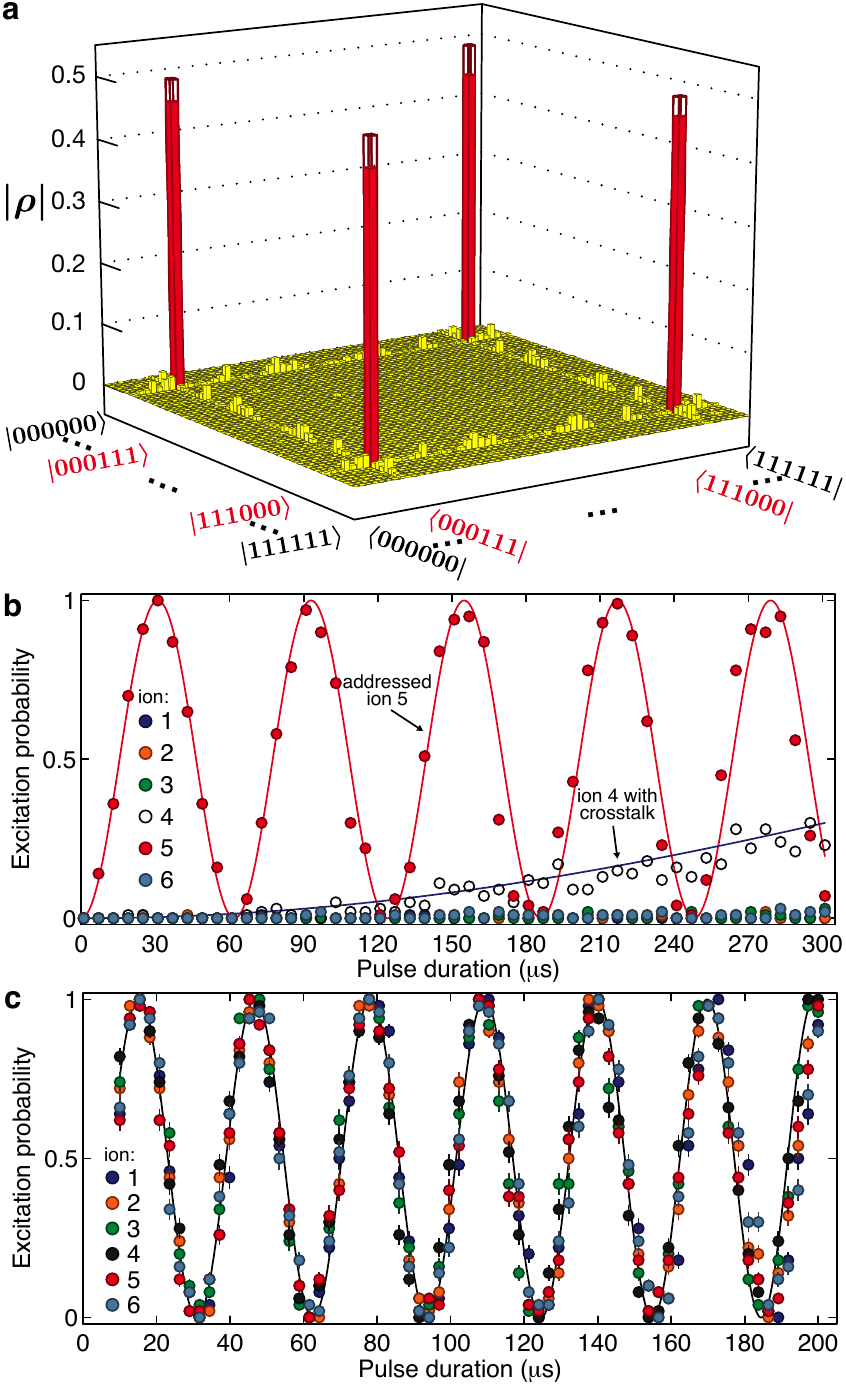}\\

\caption{\textbf{State and measurement characterization for the
    six-qubit DIEW measurement.} \textbf{a},~Reconstructed density
  matrix $\rho$ (absolute value) of six-qubit state
  $\sim|000111\rangle+e^{i\pi/2}|111000\rangle$. The reconstructed
  state has a fidelity of 91.9(3)\% with the target
  state. \textbf{b},~Single-qubit excitations while driving a
  single-qubit rotation of the fifth qubit (red markers) on a six-ion
  string, in this case, the crosstalk, due to imperfect focusing, is
  biggest for the fourth ion (black open circle markers). The red line
  corresponds to a sinusoidal fit to the excitation of the fifth ion,
  and the black line to the fourth ion. The corresponding fits to the
  excitation of ions 1, 2, 3 and 6 (markers dark blue, orange, green,
  and light blue) are not shown as the respective excitations are
  negligible. \textbf{c},~Single-qubit excitations during a six-qubit
  collective rotation. All ions are homogeneously excited as the pulse
  duration is increased. The sinusoidal fit, shown as a black line, is
  done for all points.  Error bars are 1$\sigma$ where shown,
  otherwise smaller than the marker. }
\end{figure}

In this setup, the subsystems may indeed leave the qubit subspace,
albeit with an expected very small probability, during initialization,
preparation and measurement of the entangled state {\color{nred}(see Methods)}.
Using a DIEW allows us to discard these considerations safely.

{\color{nred}In practice we observe that applying a qubit rotation on an ion produces an unintended rotation on neighboring ions (see Fig.~1b and Methods).} 
Because this
{\color{nred}crosstalk} effect is not included in the device-independent framework, which
assumes all measurements to be defined locally, we analysed the impact
that these crosstalks can have on the bound $B_{nm}$. For simplicity,
however, we only considered a crosstalk upper bound
$\epsilon$. In our system we found that
$\epsilon=\{1.6,5,4\}\%\pm\{0.1,1,1\}\%$ for $\{3,4,6\}$ ions. We then
computed numerically the maximum impact $\Delta
I^{CT}=I^\epsilon_{bisep}-I^{\epsilon=0}_{bisep}$ that crosstalks
bounded by $\epsilon$ can have on the biseparable bound for the
settings intended for the experiment ($CT$ stands for
crosstalks). Assuming that the maximum contribution of the crosstalks
to Eq.~\ref{eq:diew} is given by $\Delta I^{CT}$, we update the bound
to $B_{nm}^{CT}=B_{nm}+\Delta I^{CT}$.

With three ions, we performed a DIEW measurement on a state prepared
towards $|GHZ\rangle_3$ as described above.  For the m=2(3) settings
DIEW of Eq.~\ref{eq:setts} each measurement set consists of 8 (18)
measurements. To acquire significant statistics, we performed 6 (264)
measurement sets where each measurement was done on 250 consecutive
copies of the state (Table I). For each set the order of the
measurements was chosen randomly in order to avoid correlations
between outcomes and unaccounted changes or drifts on the experimental
conditions.

\begin{table*}[!ht]
\begin{minipage}{\textwidth}
\caption{\textbf{Summary of DIEW measurements with $\bf n=3,4,6$ qubits and
  $\bf m=2,3$ settings.}\footnote{Errors in parenthesis, 1$\sigma$, were derived
  from propagated statistics in the measured expectation values with
  1000 Monte Carlo samples.  $B$:~biseparable quantum correlations
  bound, $B^{CT}$:~updated $B$ to account for crosstalks,
  $I^{exp}$:~measured DIEW, $I^{max}$: quantum DIEW bound,
  $V$:~visibility, $q=(I^{exp}-B^{CT})/I^{exp}$:~maximum admissible
  additional white noise. The magnitude of the violation,
  $I^{exp}-B^{CT}$, is given in units of $\sigma$ (gray column). }}
\begin{ruledtabular}
{\scriptsize
\begin{tabular}{cc|ccc>{\columncolor[gray]{.9}[4pt][4pt]}cccccccc}
             $n$&$m$& $B$ &$B^{CT}$ &$I^{exp}$&$I^{exp}-B^{CT}$ ($\sigma$) &$I^{max}$ &$V=I^{exp}/I^{max}$ & $q$ (\%) & copies/meas.& meas./set&sets& copies \\\hline
3&2& 4           &4.070(4) &4.78(6)  & 12&$  4\sqrt{2}$&0.84(1)  & 15(1)  & 250& 8  & 6   &12k   \\
  &3&$6\sqrt{3}$ &10.542(9)&12.39(1) &136&$  9\sqrt{3}$&0.7948(9)& 14.9(1)& 250& 18 & 264 &1188k \\
4&2& 8           &8.43(8)  &10.42(6) & 20&$  8\sqrt{2}$&0.92(1)  & 19(1)  & 50 & 16 & 60  &48k   \\
  &3&$18\sqrt{3}$&32.5(2)  &42.53(8) &41 &$ 27\sqrt{3}$&0.909(7) & 23.7(6)& 50 & 54 & 60  &162k  \\
6&2& 32          &34.0(5)  &40(1)    &  5&$ 32\sqrt{2}$&0.88(3)  & 15(3)  & 5  & 64 & 21  &6720  \\
  &3&$162\sqrt{3}$&294(3)  &374(3)   & 19&$243\sqrt{3}$&0.89(1)  & 21(1)  & 5  & 486& 21  &51030 \\
\end{tabular}
}
\end{ruledtabular}
\end{minipage}
\end{table*}

To demonstrate genuine four- and six-partite entanglement with DIEWs,
we avoided
{\color{nred}a} decay of coherence, quadratic in the number of
qubits~\cite{monz-prl-106-130506} {\color{nred}(see Methods)}, by preparing the decoherence-free
GHZ states $\frac{1}{\sqrt{2}}(|0011\rangle+|1100\rangle)$ and
$\frac{1}{\sqrt{2}}(|000111\rangle+|111000\rangle)$. These states have
a coherence time of $\approx300$~ms, sufficient for our measurements.
For a DIEW measurement of Eq.~\ref{eq:setts}, the angles $\phi_{s_j}$
that maximize the violation for these states depend on the number of
settings $m$ and the number of qubits $n$ (see Supplementary
Information). For $m=2$ settings ($s_j=0,1$), a maximum violation is
achieved when one half of the qubits is measured with an angle
$\phi_{s_j}=s_j\frac{\pi}{2}$ and the other half with
$\phi_{s_j}'=\frac{n+1}{12n}\pi+\frac{1-s_j}{2}\pi$. For $m=3$
settings ($s_j=0,1,2$), the corresponding angles are
$\phi_{s_j}=\frac{n+1}{12n}\pi+s_j\frac{\pi}{3}$ and
$\phi_{s_j}'=\frac{2-s_j}{3}\pi$. The measured DIEW parameters shown
in Table I significantly violate the thresholds for genuine four- and
six-partite entanglement.

To compare the two- and three-settings DIEW measurements with the same
number of copies, we chose random subsets of copies from each
measurement of the above data. The results in Table II show a higher
violation for the three settings measurements with four and six
ions. This indicates that the detection of entanglement through our
witness is made easier by using more settings.  Moreover, the maximum
amount of white noise $q$ that can be mixed with the observed
statistics before loosing the violation indicates that more states are
detected by the 3-settings than the 2-settings DIEW (see also Table
I).

\begin{table}
\caption{\textbf{Comparison of DIEW measurements with $\bf m=2$ and $\bf m=3$  settings for a similar number of copies of the state of $\bf n=3,$ 4,
    and 6 qubits.}\footnote{Errors in parenthesis, 1$\sigma$, were derived from
  propagated statistics in the measured expectation values with 1000
  Monte Carlo samples.  $B^{CT}$:~updated biseparable quantum
  correlations bound to account for crosstalks, $I^{exp}$:~measured
  DIEW, $q=(I^{exp}-B^{CT})/I^{exp}$:~maximum admissible additional
  white noise. The magnitude of the violation, $I^{exp}-B^{CT}$, is
  given in units of $\sigma$ (gray column).}}
\begin{ruledtabular}
{\footnotesize
\begin{tabular}{cc|c>{\columncolor[gray]{.9}[4pt][4pt]}ccc}
$n$&$m$& $I^{exp}$ & $I^{exp}-B^{CT}$ ($\sigma$) & $q$ (\%) & copies\\\hline
3 & 2 & 4.67(9) & 7 & 13(2) & 4800\\
  & 3 & 12.2(2) & 8 & 14(2) & 4752 \\
4 & 2 & 10.5(2) & 10 & 20(2) & 3840 \\
  & 3 & 41.7(6) & 14 & 22(2) & 3240 \\
6 & 2 & 41(1)   & 6  & 17(3) & 1344 \\
  & 3 & 376(6)  & 12 & 22(2) & 1458 
\end{tabular}
}
\end{ruledtabular}
\end{table}

Because DIEW inequalities with $m=2$ settings are
equivalent to $n$-partite Svetlichny
inequalities~\cite{collins-prl-88-170405,seevinck-prl-89-060401}, they
detect not only multipartite entanglement but also multipartite quantum
nonlocality~\cite{bancal-prl-106-250404}. However,
for $m\ge3$ settings, the DIEW inequalities are not
Svetlichny inequalities: Setlichny inequalities witness both, multipartite
entanglement and quantum nonlocality, but DIEW inequalities, in
general, only witness multipartite entanglement (multipartite
entanglement is necessary in order to show multipartite quantum
nonlocality)~\cite{bancal-prl-106-250404}. Multipartite entanglement
can thus be detected device-independently even when no Svetlichny
inequality can be violated. This is indicated by the results of
Table~II for $n=6$: For some amount of additional noise, the violation
of the 2-settings inequality disappears while the 3-settings
inequality can still be violated. Note that violation of a
three-partite Svetlichny inequality was previously demonstrated with
photons in absence of crosstalks~\cite{lavoie-njp-11-073051}, but
leaving the locality and detection loopholes open.

With today's technology, the different parts of an entangled system
cannot always be well separated from each other. However, clear
separation of the parts is certainly a desired feature in the long
term, since it is necessary to take full advantage of quantum
entanglement. As shown above, entanglement can be demonstrated in
these circumstances without relying on a precise description of the
measurement devices used, a knowledge which is never available
completely. The achieved demonstration was thus more robust than
conventional entanglement demonstrations and nevertheless simple. In
order to guarantee the absence of crosstalks, we look forward to a
demonstration with clearer separation between the parties or closing
the locality loophole. On the theoretical side, further work is needed
to conceive device-independent tests of resource states sufficient for
interesting quantum simulation or communication protocols.\\

\noindent\textbf{Acknowledgments} We thank Y.-C. Liang for useful
discussions.  We gratefully acknowledge support by the Austrian
Science Fund (FWF) through the SFB FoQuS (FWF Project No. F4006-N16)
and by the Swiss NCCR “Quantum Science and Technology”, the CHIST-ERA
DIQIP, and the European ERC-AG QORE.\\



\noindent\textbf{METHODS SUMMARY}

\textbf{DIEW and tomography.} For $n$ parties and $m$ possible
measurement settings per party, the total number of measurements
settings required for the witness in Eq.~\eqref{eq:diew} is $2m^{n-1}$
whereas tomography requires $(d^2-1)^n$ measurements, where $d$ is the
dimension of each party or subsystem~\cite{guhne-physrep-474-1}. For
$m=3$ and qubits ($d=2$), for example, this witness requires two
thirds the number of measurements required for full quantum state
tomography (this advantage becomes more evident as the dimension of
the system increases). In addition, the required post-processing is
trivial in comparison with state reconstruction methods. Naturally,
considering $m=2$ gives the minimum total number of measurements
required, but as shown in the main text, considering $m=3$ should
result in more robust violations. For $m=4$ and $d=2$, however, the
DIEW measurement requires more measurements than tomography.

\textbf{Experimental leak of qubit subspace.} The qubits are
initialized into the state $4S_{1/2}(m_j=-1/2)=|1\rangle$ with a
99.9\% probability, with the remaining population in the
$4S_{1/2}(m_j=+1/2)$. Also, during the entangling operation, the
motional ground state can be unintentionally abandoned for higher
motional states. Furthermore, our state detection protocol identifies
as population in the state $|1\rangle$ any observation of fluorescence
on the dipole transition, including $4S_{1/2}(m_j=-1/2)$ and higher
vibrational states.

\textbf{Experimental crosstalks.} In practice we observe that applying
the rotation $\exp(-i\frac{\varphi_j}{2}\sigma_z^{(j)})$ on ion $j$
produces an unintended rotation $\varphi_k$ on qubit $k$ because of
residual light from imperfect
focusing~\cite{schmidt-kaler-apblo-77-789}(e.g., Fig.~1b). This
imperfection affects the net rotation applied on ion $k$, so that the
measurement applied on ion $k$ depends slightly on the one applied on
ion $j$. These crosstalks can be quantified by the matrix
$C_{jk}=\varphi_{k}/\varphi_{j}$, and the operation intended only on
ion $j$ is then described by the operation on all ions
$\exp(-i\sum_k\frac{\varphi_j}{2}C_{jk}\sigma^{(k)}_z)$. For
simplicity, we only considered a crosstalk upper bound
$\epsilon>C_{jk}$.

\textbf{Collective decay of three-qubit state.} Although the measured
DIEW parameters shown in Table I for $n=3$ parties do violate the
threshold for genuine three-partite entanglement, the observed
visibilities of about 80\% are incompatible with the 97\% fidelity
reported earlier~\cite{monz-prl-106-130506}. This disagreement is due
to the collective decay of the state during the DIEW
measurement. Although the single-qubit coherence time in this
experiment was $\approx$10~ms, the coherence time of the state
$|GHZ\rangle_3$ was only $\approx$2~ms. Because each single-qubit
rotation took $\approx100~\mu s$ for a $2\pi$ rotation (for collective
rotations $\approx20~\mu s/2\pi$), and for each measurement the qubits
were rotated sequentially, the state was significantly decaying during
the DIEW measurement.

\bibliographystyle{naturemag}
\bibliography{all}

\begin{thebibliography}{10}
\expandafter\ifx\csname url\endcsname\relax
  \def\url#1{\texttt{#1}}\fi
\expandafter\ifx\csname urlprefix\endcsname\relax\def\urlprefix{URL }\fi
\providecommand{\bibinfo}[2]{#2}
\providecommand{\eprint}[2][]{\url{#2}}

\bibitem{guhne-physrep-474-1}
\bibinfo{author}{G{\"u}hne, O.} \& \bibinfo{author}{T{\'o}th, G.}
\newblock \bibinfo{title}{{Entanglement detection}}.
\newblock \emph{\bibinfo{journal}{Phys. Rep.}} \textbf{\bibinfo{volume}{474}},
  \bibinfo{pages}{1--75} (\bibinfo{year}{2009}).

\bibitem{rosset-arxiv-1203.0911}
\bibinfo{author}{Rosset, D.}, \bibinfo{author}{Ferretti-Sch{\"o}bitz, R.},
  \bibinfo{author}{Bancal, J.-D.}, \bibinfo{author}{Gisin, N.} \&
  \bibinfo{author}{Liang, Y.-C.}
\newblock \bibinfo{title}{{Imperfect measurements settings: implications on
  quantum state tomography and entanglement witnesses}}.
\newblock \emph{\bibinfo{journal}{arXiv:12030911}}  (\bibinfo{year}{2012}).

\bibitem{bancal-prl-106-250404}
\bibinfo{author}{Bancal, J.-D.}, \bibinfo{author}{Gisin, N.},
  \bibinfo{author}{Liang, Y.-C.} \& \bibinfo{author}{Pironio, S.}
\newblock \bibinfo{title}{{Device-Independent Witnesses of Genuine Multipartite
  Entanglement}}.
\newblock \emph{\bibinfo{journal}{Phys. Rev. Lett.}}
  \textbf{\bibinfo{volume}{106}}, \bibinfo{pages}{250404}
  (\bibinfo{year}{2011}).

\bibitem{briegel-prl-81-5932}
\bibinfo{author}{Briegel, H.~J.}, \bibinfo{author}{D{\"u}r, W.},
  \bibinfo{author}{Cirac, J.} \& \bibinfo{author}{Zoller, P.}
\newblock \bibinfo{title}{{Quantum Repeaters: The Role of Imperfect Local
  Operations in Quantum Communication}}.
\newblock \emph{\bibinfo{journal}{Phys. Rev. Lett.}}
  \textbf{\bibinfo{volume}{81}}, \bibinfo{pages}{5932--5935}
  (\bibinfo{year}{1998}).

\bibitem{ekert-prl-67-661}
\bibinfo{author}{Ekert, A.~K.}
\newblock \bibinfo{title}{{Quantum cryptography based on Bell's theorem}}.
\newblock \emph{\bibinfo{journal}{Phys. Rev. Lett.}}
  \textbf{\bibinfo{volume}{67}}, \bibinfo{pages}{661--663}
  (\bibinfo{year}{1991}).

\bibitem{gisin-rmp-74-145}
\bibinfo{author}{Gisin, N.}, \bibinfo{author}{Ribordy, G.},
  \bibinfo{author}{Tittel, W.} \& \bibinfo{author}{Zbinden, H.}
\newblock \bibinfo{title}{{Quantum cryptography}}.
\newblock \emph{\bibinfo{journal}{Rev. Mod. Phys.}}
  \textbf{\bibinfo{volume}{74}}, \bibinfo{pages}{145--195}
  (\bibinfo{year}{2002}).

\bibitem{briegel-natphys-5-19}
\bibinfo{author}{Briegel, H.~J.}, \bibinfo{author}{Browne, D.~E.},
  \bibinfo{author}{D{\"u}r, W.}, \bibinfo{author}{Raussendorf, R.} \&
  \bibinfo{author}{Nest, M. V.~d.}
\newblock \bibinfo{title}{{Measurement-based quantum computation}}.
\newblock \emph{\bibinfo{journal}{Nature Phys.}} \textbf{\bibinfo{volume}{5}},
  \bibinfo{pages}{19} (\bibinfo{year}{2009}).

\bibitem{dicarlo-nature-467-574}
\bibinfo{author}{DiCarlo, L.} \emph{et~al.}
\newblock \bibinfo{title}{{Preparation and measurement of three-qubit
  entanglement in a superconducting circuit}}.
\newblock \emph{\bibinfo{journal}{Nature}} \textbf{\bibinfo{volume}{467}},
  \bibinfo{pages}{574--578} (\bibinfo{year}{2010}).

\bibitem{neumann-science-320-1326}
\bibinfo{author}{Neumann, P.} \emph{et~al.}
\newblock \bibinfo{title}{{Multipartite Entanglement Among Single Spins in
  Diamond}}.
\newblock \emph{\bibinfo{journal}{Science}} \textbf{\bibinfo{volume}{320}},
  \bibinfo{pages}{1326--1329} (\bibinfo{year}{2008}).

\bibitem{saffman-rmp-82-2313}
\bibinfo{author}{Saffman, M.}, \bibinfo{author}{Walker, T.~G.} \&
  \bibinfo{author}{M{\o}lmer, K.}
\newblock \bibinfo{title}{{Quantum information with Rydberg atoms}}.
\newblock \emph{\bibinfo{journal}{Rev. Mod. Phys.}}
  \textbf{\bibinfo{volume}{82}}, \bibinfo{pages}{2313--2363}
  (\bibinfo{year}{2010}).

\bibitem{blatt-nature-453-1008}
\bibinfo{author}{Blatt, R.} \& \bibinfo{author}{Wineland, D.}
\newblock \bibinfo{title}{{Entangled states of trapped atomic ions}}.
\newblock \emph{\bibinfo{journal}{Nature}} \textbf{\bibinfo{volume}{453}},
  \bibinfo{pages}{1008--1015} (\bibinfo{year}{2008}).

\bibitem{yao-natphot-6-225}
\bibinfo{author}{Yao, X.-C.} \emph{et~al.}
\newblock \bibinfo{title}{{Observation of eight-photon entanglement}}.
\newblock \emph{\bibinfo{journal}{Nature Photon.}}
  \textbf{\bibinfo{volume}{6}}, \bibinfo{pages}{225--228}
  (\bibinfo{year}{2012}).

\bibitem{gao-natphys-6-331}
\bibinfo{author}{Gao, W.-B.} \emph{et~al.}
\newblock \bibinfo{title}{{Experimental demonstration of a hyper-entangled
  ten-qubit Schr{\"o}dinger cat state}}.
\newblock \emph{\bibinfo{journal}{Nature Phys.}} \textbf{\bibinfo{volume}{6}},
  \bibinfo{pages}{331--335} (\bibinfo{year}{2010}).

\bibitem{monz-prl-106-130506}
\bibinfo{author}{Monz, T.} \emph{et~al.}
\newblock \bibinfo{title}{{14-Qubit Entanglement: Creation and Coherence}}.
\newblock \emph{\bibinfo{journal}{Phys. Rev. Lett.}}
  \textbf{\bibinfo{volume}{106}}, \bibinfo{pages}{130506}
  (\bibinfo{year}{2011}).

\bibitem{semenov-pra-83-032119}
\bibinfo{author}{Semenov, A.} \& \bibinfo{author}{Vogel, W.}
\newblock \bibinfo{title}{{Fake violations of the quantum Bell-parameter
  bound}}.
\newblock \emph{\bibinfo{journal}{Phys. Rev. A}} \textbf{\bibinfo{volume}{83}},
  \bibinfo{pages}{032119} (\bibinfo{year}{2011}).

\bibitem{altepeter-lnp-649-113}
\bibinfo{author}{Altepeter, J.~B.}, \bibinfo{author}{James, D. F.~V.} \&
  \bibinfo{author}{Kwiat, P.~G.}
\newblock \bibinfo{title}{{Qubit quantum state tomography}}.
\newblock \emph{\bibinfo{journal}{Lecture Notes of Physics}}
  \textbf{\bibinfo{volume}{649}}, \bibinfo{pages}{113--145}
  (\bibinfo{year}{2004}).

\bibitem{lvovsky-rmp-81-299}
\bibinfo{author}{Lvovsky, A.~I.}
\newblock \bibinfo{title}{{Continuous-variable optical quantum-state
  tomography}}.
\newblock \emph{\bibinfo{journal}{Rev. Mod. Phys.}}
  \textbf{\bibinfo{volume}{81}}, \bibinfo{pages}{299--332}
  (\bibinfo{year}{2009}).

\bibitem{audenaert-njp-11-023028}
\bibinfo{author}{Audenaert, K.} \& \bibinfo{author}{Scheel, S.}
\newblock \bibinfo{title}{{Quantum tomographic reconstruction with error bars:
  a Kalman filter approach}}.
\newblock \emph{\bibinfo{journal}{New J. Phys.}} \textbf{\bibinfo{volume}{11}},
  \bibinfo{pages}{023028} (\bibinfo{year}{2009}).

\bibitem{acin-prl-97-120405}
\bibinfo{author}{Ac{\'\i}n, A.}, \bibinfo{author}{Gisin, N.} \&
  \bibinfo{author}{Masanes, L.}
\newblock \bibinfo{title}{{From Bell's Theorem to Secure Quantum Key
  Distribution}}.
\newblock \emph{\bibinfo{journal}{Phys. Rev. Lett.}}
  \textbf{\bibinfo{volume}{97}}, \bibinfo{pages}{120405}
  (\bibinfo{year}{2006}).

\bibitem{gallego-prl-105-230501}
\bibinfo{author}{Gallego, R.}, \bibinfo{author}{Brunner, N.},
  \bibinfo{author}{Hadley, C.} \& \bibinfo{author}{Ac{\'\i}n, A.}
\newblock \bibinfo{title}{{Device-Independent Tests of Classical and Quantum
  Dimensions}}.
\newblock \emph{\bibinfo{journal}{Phys. Rev. Lett.}}
  \textbf{\bibinfo{volume}{105}}, \bibinfo{pages}{230501}
  (\bibinfo{year}{2010}).

\bibitem{hendrych-natphys-8-588}
\bibinfo{author}{Hendrych, M.} \emph{et~al.}
\newblock \bibinfo{title}{{Experimental estimation of the dimension of
  classical and quantum systems}}.
\newblock \emph{\bibinfo{journal}{Nature Phys.}} \textbf{\bibinfo{volume}{8}},
  \bibinfo{pages}{588--591} (\bibinfo{year}{2012}).

\bibitem{ahrens-natphys-8-592}
\bibinfo{author}{Ahrens, J.}, \bibinfo{author}{Badziag, P.},
  \bibinfo{author}{Cabello, A.} \& \bibinfo{author}{Bourennane, M.}
\newblock \bibinfo{title}{{Experimental device-independent tests of classical
  and quantum dimensions}}.
\newblock \emph{\bibinfo{journal}{Nature Phys.}} \textbf{\bibinfo{volume}{8}},
  \bibinfo{pages}{592--595} (\bibinfo{year}{2012}).

\bibitem{pal-pra-83-062123}
\bibinfo{author}{P{\'a}l, K.} \& \bibinfo{author}{V{\'e}rtesi, T.}
\newblock \bibinfo{title}{{Multisetting Bell-type inequalities for detecting
  genuine multipartite entanglement}}.
\newblock \emph{\bibinfo{journal}{Phys. Rev. A}} \textbf{\bibinfo{volume}{83}},
  \bibinfo{pages}{062123} (\bibinfo{year}{2011}).

\bibitem{bancal-jphysa-45-125301}
\bibinfo{author}{Bancal, J.-D.}, \bibinfo{author}{Branciard, C.},
  \bibinfo{author}{Brunner, N.}, \bibinfo{author}{Gisin, N.} \&
  \bibinfo{author}{Liang, Y.-C.}
\newblock \bibinfo{title}{{A framework for the study of symmetric
  full-correlation Bell-like inequalities}}.
\newblock \emph{\bibinfo{journal}{J. Phys. A: Math. Theor.}}
  \textbf{\bibinfo{volume}{45}}, \bibinfo{pages}{125301}
  (\bibinfo{year}{2012}).

\bibitem{schmidt-kaler-apblo-77-789}
\bibinfo{author}{Schmidt-Kaler, F.} \emph{et~al.}
\newblock \bibinfo{title}{{How to realize a universal quantum gate with trapped
  ions}}.
\newblock \emph{\bibinfo{journal}{Appl. Phys. B: Lasers Opt.}}
  \textbf{\bibinfo{volume}{77}}, \bibinfo{pages}{789--796}
  (\bibinfo{year}{2003}).

\bibitem{molmer-prl-82-1835}
\bibinfo{author}{M{\o}lmer, K.} \& \bibinfo{author}{S{\o}rensen, A.}
\newblock \bibinfo{title}{{Multiparticle Entanglement of Hot Trapped Ions}}.
\newblock \emph{\bibinfo{journal}{Phys. Rev. Lett.}}
  \textbf{\bibinfo{volume}{82}}, \bibinfo{pages}{1835--1838}
  (\bibinfo{year}{1999}).

\bibitem{sackett-nat-404-256}
\bibinfo{author}{Sackett, C.~A.} \emph{et~al.}
\newblock \bibinfo{title}{{Experimental entanglement of four particles}}.
\newblock \emph{\bibinfo{journal}{Nature}} \textbf{\bibinfo{volume}{404}},
  \bibinfo{pages}{256--259} (\bibinfo{year}{2000}).

\bibitem{collins-prl-88-170405}
\bibinfo{author}{Collins, D.}, \bibinfo{author}{Gisin, N.},
  \bibinfo{author}{Popescu, S.}, \bibinfo{author}{Roberts, D.} \&
  \bibinfo{author}{Scarani, V.}
\newblock \bibinfo{title}{{Bell-Type Inequalities to Detect True n-Body
  Nonseparability}}.
\newblock \emph{\bibinfo{journal}{Phys. Rev. Lett.}}
  \textbf{\bibinfo{volume}{88}}, \bibinfo{pages}{170405}
  (\bibinfo{year}{2002}).

\bibitem{seevinck-prl-89-060401}
\bibinfo{author}{Seevinck, M.} \& \bibinfo{author}{Svetlichny, G.}
\newblock \bibinfo{title}{{Bell-Type Inequalities for Partial Separability in
  N-Particle Systems and Quantum Mechanical Violations}}.
\newblock \emph{\bibinfo{journal}{Phys. Rev. Lett.}}
  \textbf{\bibinfo{volume}{89}}, \bibinfo{pages}{060401}
  (\bibinfo{year}{2002}).

\bibitem{lavoie-njp-11-073051}
\bibinfo{author}{Lavoie, J.}, \bibinfo{author}{Kaltenbaek, R.} \&
  \bibinfo{author}{Resch, K.~J.}
\newblock \bibinfo{title}{{Experimental violation of Svetlichny's inequality}}.
\newblock \emph{\bibinfo{journal}{New J. Phys.}} \textbf{\bibinfo{volume}{11}},
  \bibinfo{pages}{073051} (\bibinfo{year}{2009}).

\end{thebibliography}


\clearpage


\section*{Supplementary material}

# Device independent demonstration of genuine multipartite entanglement
# — SUPPLEMENTARY MATERIAL —

Julio T. Barreiro[1*], Jean-Daniel Bancal[2*], Philipp Schindler[1], Daniel Nigg[1],
Markus Hennrich[1], Thomas Monz[1], Nicolas Gisin[2], Rainer Blatt[1,3]

[1]Institut für Experimentalphysik, Universität Innsbruck, Technikerstr. 25, A-6020 Innsbruck, Austria
[2]Group of Applied Physics, University of Geneva, Geneva, Switzerland
[3]Institut für Quantenoptik und Quanteninformation, Österreichische Akademie der Wissenschaften,
   Technikerstr. 21A, A-6020 Innsbruck, Austria
* These authors contributed equally to this work.

**CONTENTS**

## I. EXPLICIT EXAMPLE FOR THE EVALUATION OF THE DIEW PARAMETER $I_{nm}$

Here we show an explicit example of the terms involved on the DIEW parameter

$$I_{nm} = \sum_{s \equiv 0 \ (\text{mod } m)} (-1)^{s/m} E_{\vec{s}} + \sum_{s \equiv 1 \ (\text{mod } m)} (-1)^{(s-1)/m} E_{\vec{s}} \quad (1)$$

where $E_{\vec{s}} = \sum_{\vec{r}}(-1)^r P(\vec{r}|\vec{s})$ is the $n$-partite correlator; $s = \sum_j s_j$ and $r = \sum_j r_j$ are the sums over all indices used in the measurement vector $\vec{s}$ and the outcome vector $\vec{r}$ (the colors will be evident in a moment). Obviously, for any settings $\vec{s}$ the following is true $\sum_{\vec{r}} P(\vec{r}|\vec{s}) = 1$.

For $n = 3$ qubits and $m = 3$ measurement settings, there are $3 \times 3 \times 3 = 27$ possible measurements as shown in the leftmost column in Table I. However, because the parameter $I_{n=3,m=3}$ involves only those settings for which $s$ mod 3 is equal to 0 or 1, only 18 of the 27 measurement settings are necessary. These entries, involving $s$ modulo 3 equal to 0 or 1, are shown in red and green font colors in Table I.

For our scenario where a measurement on each party has one of two possible outcomes, in our example with three parties, there are then eight possibilities, that is the vector of outcomes can take any of the following values:

$$\vec{r} \in \{\mathbf{000}, \mathbf{001}, \mathbf{010}, \mathbf{011}, \mathbf{100}, \mathbf{101}, \mathbf{110}, \mathbf{111}\},$$

where the vectors of possible outcomes are written here in boldface to distinguish them from the vector of measurement settings.

For example, when the vector of measurement settings is $\vec{s} = (s_1, s_2, s_3) = (0, 1, 2)$, then $s = 3$ and $s \equiv 0$ mod 3, as shown in the row colored light gray in Table I. The 3-partite correlator term $E_{012} = \sum_{\vec{r}}(-1)^r P(\vec{r}|012)$ is therefore involved in the parameter $I_{33}$ as

$$(-1)^{(0+1+2)/3} E_{012} = -\big[P(\mathbf{000}|012) - P(\mathbf{001}|012)$$
$$+ P(\mathbf{010}|012) - P(\mathbf{011}|012)$$
$$+ P(\mathbf{100}|012) - P(\mathbf{101}|012)$$
$$+ P(\mathbf{110}|012) - P(\mathbf{111}|012)\big]$$

In the scenario considered in which the measure-

TABLE I Measurement settings considered in the evaluation of the DIEW parameter $I_{33}$, see text.

| $s_1$ | $s_2$ | $s_3$ | $s$ | $s$ mod 3 | $(-1)^{s/3}$ | $(-1)^{(s-1)/3}$ |
|---|---|---|---|---|---|---|
| 0 | 0 | 0 | 0 | 0 | 1 | |
| 0 | 0 | 1 | 1 | 1 | | 1 |
| 0 | 0 | 2 | 2 | 2 | | |
| 0 | 1 | 0 | 1 | 1 | | 1 |
| 0 | 1 | 1 | 2 | 2 | | |
| 0 | 1 | 2 | 3 | 0 | -1 | |
| 0 | 2 | 0 | 2 | 2 | | |
| 0 | 2 | 1 | 3 | 0 | -1 | |
| 0 | 2 | 2 | 4 | 1 | | -1 |
| 1 | 0 | 0 | 1 | 1 | | 1 |
| 1 | 0 | 1 | 2 | 2 | | |
| 1 | 0 | 2 | 3 | 0 | -1 | |
| 1 | 1 | 0 | 2 | 2 | | |
| 1 | 1 | 1 | 3 | 0 | -1 | |
| 1 | 1 | 2 | 4 | 1 | | -1 |
| 1 | 2 | 0 | 3 | 0 | -1 | |
| 1 | 2 | 1 | 4 | 1 | | -1 |
| 1 | 2 | 2 | 5 | 2 | | |
| 2 | 0 | 0 | 2 | 2 | | |
| 2 | 0 | 1 | 3 | 0 | -1 | |
| 2 | 0 | 2 | 4 | 1 | | -1 |
| 2 | 1 | 0 | 3 | 0 | -1 | |
| 2 | 1 | 1 | 4 | 1 | | -1 |
| 2 | 1 | 2 | 5 | 2 | | |
| 2 | 2 | 0 | 4 | 1 | | -1 |
| 2 | 2 | 1 | 5 | 2 | | |
| 2 | 2 | 2 | 6 | 0 | 1 | |

ments are prescribed by angles, the settings are $\{\phi_{s_1}, \phi_{s_2}, \phi_{s_3}\} = \{\phi_0, \phi_1, \phi_2\}$.

Another example, when $\vec{s} = (s_1, s_2, s_3) = (2, 1, 1)$, then $s = 4$ and $s \equiv 1 \mod 3$, as shown in the row colored dark gray in Table I. The 3-partite correlator term $E_{211} = \sum_{\vec{r}}(-1)^r P(\vec{r}|211)$ is therefore involved in the parameter $I_{33}$ as

$$(-1)^{((2+1+1)-1)/3} E_{211} = -[P(\mathbf{000}|211) - P(\mathbf{001}|211)\\
+ P(\mathbf{010}|211) - P(\mathbf{011}|211)\\
+ P(\mathbf{100}|211) - P(\mathbf{101}|211)\\
+ P(\mathbf{110}|211) - P(\mathbf{111}|211)]$$

Again, in the scenario considered in which the measurements are prescribed by angles, the settings are $\{\phi_{s_1}, \phi_{s_2}, \phi_{s_3}\} = \{\phi_2, \phi_1, \phi_1\}$. For the 3-partite GHZ state, for example, the optimum angles calculated as described below, are

$$\phi_0 = -\frac{\pi}{18}$$
$$\phi_1 = -\frac{\pi}{18} + \frac{\pi}{3}$$
$$\phi_2 = -\frac{\pi}{18} + \frac{2\pi}{3}.$$

## II. THE DIEW PARAMETER AS THE EXPECTATION VALUE OF THE QUANTUM OPERATOR $\mathcal{I}_{nm}$

The value of $I_{nm}$ can be understood as the expectation value of the operator $\mathcal{I}_{nm}$ on the measured quantum state $\rho$, that is $I_{nm} = \text{Tr}(\mathcal{I}_{nm}\rho)$. From the definition of the parameter $I_{nm}$ and the observed correlations $P(\vec{r}|\vec{s}) = \text{Tr}\left[\bigotimes_{j=1}^n M^j_{r_j|s_j} \rho\right]$, the operator can naturally be written as

$$\mathcal{I}_{nm} = \sum_{s \equiv 0 \pmod{m}} (-1)^{s/m} \sum_{\vec{r}} (-1)^r \bigotimes_{j=1}^n M^j_{r_j|s_j} +\\
\sum_{s \equiv 1 \pmod{m}} (-1)^{(s-1)/m} \sum_{\vec{r}} (-1)^r \bigotimes_{j=1}^n M^j_{r_j|s_j}\\
= \sum_{s \equiv 0 \pmod{m}} \sum_{\vec{r}} (-1)^{r+s/m} \bigotimes_{j=1}^n M^j_{r_j|s_j} +\\
\sum_{s \equiv 1 \pmod{m}} \sum_{\vec{r}} (-1)^{r+(s-1)/m} \bigotimes_{j=1}^n M^j_{r_j|s_j}$$

where $M^j_{r_j|s_j}$ is the measurement operator acting locally on subsystem $j$ corresponding to setting $s_j$ and outcome $r_j$.

## III. OPTIMAL MEASUREMENT SETTINGS FOR THE DIEW PARAMETER $I_{nm}$

For every quantum state $\rho$, the measurement settings that yield the largest value of $I_{nm}$ can be determined by numerical optimization. In general, this can be achieved by parametrizing the measurement settings $M^j_{r_j|s_j}$ of each party $j$ inside their respective Hilbert space $\mathcal{H}^j$ with a finite number of real variables. Gathering these variables for the measurement settings of all parties in a vector of parameters $\vec{\alpha}$ then allows one to express the value of $I_{nm}$ as a function of $\vec{\alpha}$ only, by letting $P(\vec{r}|\vec{s}) = Tr(\otimes_{j=1}^n M^j_{r_j|s_j}(\vec{\alpha}) \rho)$ in equation (1). The optimal settings are then given by $M^j_{r_j|s_j}(\vec{\alpha}^*)$ where $\vec{\alpha}^* = \arg\max I_{nm}(\vec{\alpha})$.

In the case that the tripartite expression with $m = 3$ settings $I_{3,3}$ is evaluated on a GHZ state

$$|GHZ\rangle = \frac{1}{\sqrt{2}}(|000\rangle + |111\rangle),$$

for instance, the measurement operators corresponding to setting $s_j$ of party $j$ can be written as:

$$M^j_{0|s_j} = \frac{\mathbb{1} + \vec{n} \cdot \vec{\sigma}}{2}, \quad M^j_{1|s_j} = \mathbb{1} - M^j_{0|s_j}, \quad (2)$$

where $||\vec{n}|| = 1$ and $\vec{\sigma} = (\sigma_x, \sigma_y, \sigma_z)$ is the vector of Pauli matrices. They can thus be parametrized by two real numbers $\vartheta^j_{s_j}$ and $\varphi^j_{s_j}$ by letting:

$$\vec{n} = \begin{pmatrix} \sin(\vartheta^j_{s_j})\cos(\varphi^j_{s_j}) \\ \sin(\vartheta^j_{s_j})\sin(\varphi^j_{s_j}) \\ \cos(\vartheta^j_{s_j}) \end{pmatrix}. \quad (3)$$

18 parameters are thus enough to parametrize all measurements appearing in the $I_{3,3}$ expression. These can be gathered in the variable $\vec{\alpha} = \{\vartheta^j_{r_j|s_j}, \varphi^j_{r_j|s_j}\}_{\vec{s},j} = \{\vartheta^1_0, \vartheta^1_1, \vartheta^1_2, \vartheta^2_0, \ldots \vartheta^2_2, \ldots, \vartheta^3_2, \varphi^1_0, \ldots, \varphi^3_2\}$. The value of $I_{3,3}$ can then be expressed as a function of $\vec{\alpha}$ directly, by substituting $P(\vec{r}|\vec{s}) = \langle GHZ|M^1_{r_1|s_1} \otimes M^2_{r_2|s_2} \otimes M^3_{r_3|s_3}|GHZ\rangle$ in equation (1). For instance, the probability of observing outcomes $\mathbf{000}$ when all parties use their first setting is given by

$$P(\mathbf{000}|000) = \langle GHZ|M^1_{\mathbf{0}|0} \otimes M^2_{\mathbf{0}|0} \otimes M^3_{\mathbf{0}|0}|GHZ\rangle \quad (4)$$

After optimization of $I_{3,3}$ over $\vec{\alpha}$, one finds that all values of $\vartheta^j_{s_j}$ take the value $\pi/2$. This means that the settings yielding the largest value of $I_{3,3}$ upon measurement of a GHZ state lie in the x-y plane of the Bloch sphere.



\end{document}